\RequirePackage{fix-cm}
\documentclass[smallextended]{svjour3}       
\smartqed  
\usepackage{graphicx}
\begin{document}

\voffset 1in

\title{Syndrome Measurement Order for the [[7,1,3]] Quantum Error Correction Code
}

\titlerunning{Syndrome Measurement Order}        

\author{Yaakov S. Weinstein}


\institute{Quantum Information Science Group, {\sc Mitre}, 200 Forrestal Rd., Princeton, NJ 08540, USA
              \email{weinstein@mitre.org}           
}

\date{Received: date / Accepted: date}

\maketitle

\begin{abstract}
In this work we explore the accuracy of quantum error correction depending of the order of the implemented syndrome measurements. CSS codes require bit-flip and phase flip-syndromes be measured separately. To comply with fault tolerant demands and to maximize accuracy this set of syndrome measurements should be repeated allowing for flexibility in the order of their implementation. We examine different possible orders of Shor state and Steane state syndrome measurements for the [[7,1,3]] quantum error correction code. We find that the best choice of syndrome order, determined by the fidelity of the state after noisy error correction, will depend on the error environment. We also compare the fidelity when syndrome measurements are done with Shor states versus Steane states and find that Steane states generally, but not always, lead to final states with higher fidelity. Together, these results allow a quantum computer programmer to choose the optimal syndrome measurement scheme based on the system's error environment.   
\keywords{quantum error correction \and syndrome measurements \and  [[7,1,3]] code}
\PACS{03.67.Pp \and 03.67.-a \and 03.67.Lx}
\end{abstract}

If quantum computation is to become a reality there must be robust quantum error correction (QEC) codes \cite{book,ShorQEC,CSS}. QEC codes encode a given number of `logical' qubits of quantum information into a larger number of physical or `data' qubits. Should one of the  physical qubits undergo an error after encoding, measurements of the parity between specified physical qubits, known as syndrome measurements (SM), will locate and identify the error. A recovery operation will then perform the appropriate correction.

QEC forms the backbone of quantum fault tolerance (QFT), a comprehensive framework that allows for successful quantum computation despite errors in basic computational components \cite{Preskill,ShorQFT,G,AGP}. The overarching concept behind QFT is to implement quantum operations within the QEC code space in such a way as to ensure that any error on a single physical qubit will remain localized, {\emph{i.~e.~}}the error will not spread to multiple qubits. This can be done by properly designing quantum protocols with the possible use of ancilla qubits, and the repetition of certain protocols in order to relegate errors to second order in error probability. Any remaining single-qubit errors can then be identified via SM and corrected by applying the appropriate recovery operation. By concatenating QEC codes the probability of logical qubit errors can be suppressed to increasingly higher orders of the physical error probability. 

A paradigmatic example of a QEC code is the [[7,1,3]] code or Steane code \cite{Steane} which stores one logical qubit in seven data qubits. As with all CSS codes, separate syndrome measurements are needed to locate and identify bit flip errors and phase flip errors. Here, we explore two methods that are compatible with the rules of QFT: the Shor state method \cite{ShorQFT} and the Steane state method \cite{Steane}. 

\begin{figure}
\begin{center}
\includegraphics[width=8.5cm]{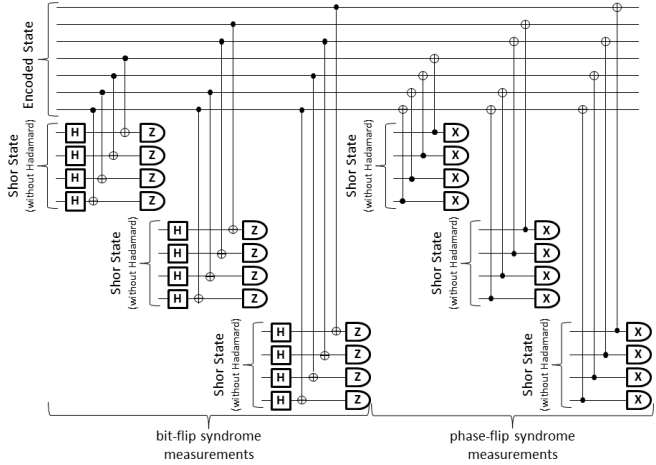}
\caption{
Fault tolerant bit-flip and phase-flip syndrome measurements for the [[7,1,3]] code using Shor states (here the Shor states are assumed to have not had the Hadamard gates applied). CNOT gates are represented by ($\bullet$) on the control qubit and ($\oplus$) on the target qubit connected by a vertical line. $H$ represents a Hadamard gate. The error syndrome is determined from the parity of the measurement outcomes of the Shor state ancilla qubits. To achieve fault tolerance the sets of syndrome measurements are repeated until the same syndrome is read out twice.}
\label{ShorSM}
\end{center}
\end{figure}

Shor states are simply GHZ states with a Hadamard applied to each qubit. To completely read out the bit flip or phase flip syndrome of the [[7,1,3]] code requires three four-qubit Shor states. Controlled-NOT (CNOT) gates are performed between specified data qubits and the qubits of the Shor state as shown in Fig.~\ref{ShorSM}. The four Shor state qubits are measured and the parity of the measurements determines the SM. Note that in consonance with the rules of QFT no Shor state qubit ever interacts with more than one data qubit thus stemming the spread of any possible error. In addition, we must ensure that no errors occurred during the SM. To do this QFT requires that the set of bit-flip and phase-flip SM be repeated until the same results are obtained twice. In our simulations we will assume that the SM sets give the same readout the first two times. 

Shor state construction is implemented via a series of CNOT gates. Verification via a parity check between two of the qubits, is implemented to ensure no errors occured during construction \cite{ShorQFT,WB}. 

\begin{figure}
\begin{center}
\includegraphics[width=5cm]{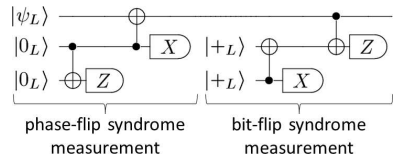}
\caption{Circuit for fault tolerant syndrome measurements for the [[7,1,3]] QEC code using Steane ancilla. Each line represents the seven physical qubits. The circuit shows that each ancilla is verified to check for errors that may have occurred during ancilla construction. }
\label{SteaneSM}
\end{center}
\end{figure}

Steane states are logical $|0\rangle$ or $|+\rangle$ states of the [[7,1,3]] QEC code used to read out phase-flip and bit-flip syndromes respectively. The SM are performed by implementing a logical CNOT gate between the data qubits and the qubits of the Steane state followed by measurement of the Steane state qubits as shown in Fig.~\ref{SteaneSM}. Unlike Shor state SM, Steane state SM need not be repeated in order to achieve fault tolerance \cite{AGP}. Nevertheless, repeating the SM does increase the fidelity of the encoded state \cite{YSWSynds}.

Steane states are constructed via the encoding gate sequence first developed in \cite{Steane}. To ensure no errors occurred during constuction two $|0\rangle$ or $|+\rangle$ states are thus constructed. Verification is done by performing a logical CNOT between the Steane states and measuring the qubits of one of the states. 

Thus, both Shor state and Steane state SM benefit from multiple implementations of the set bit-flip and phase-flip syndrome measurements. However within the set either the bit-flip or phase flip SM must be implemented first. Assuming the SM are repeated twice (which is the minimum) four, different SM orders are possible: $XZXZ$, $XZZX$, $ZXXZ$, $ZXZX$, where $X$ ($Z$) refers to the complete bit-flip (phase-flip) syndrome measurements within a set. The goal of this paper is to determine which order is optimum in various error environments. Our simulations are limited to the [[7,1,3]] QEC code. Nonetheless, we believe that our demonstration highlights the flexibility of SM order which may be an important issue for other CSS codes. 

To explore the accuracy of different SM orders we simulate the implementation of 50 single logical qubit gates with SM applied after each one. The simulations assume a nonequiprobable Pauli operator error environment \cite{QCC}. This model is a stochastic version of a biased noise model that can be formulated in terms of Hamiltonians coupling the system to an environment \cite{AP}. In our model different error types occur with different probabilities: $\sigma_x^j$ errors occur with probability $p_x$, $\sigma_y^j$ errors with probability $p_y$, and a $\sigma_z^j$ errors with probability $p_z$, for Pauli operators $\sigma_i^j$, $i = x,y,z$ on qubit $j$. When the three error probabililties are equivalent we are in a depolarizing environment. We assume that qubits taking part in a gate operation, initialization or measurement (including all the gates and measurments of the Shor state and Steane state construction and verification) will be subject to error and that errors are completely uncorrelated. 

In our simulations we utilize a sequence of 50 gates each followed by SM. The gates are represented by the composite gates $A = HST$ and $B = HT$, where $H$ is the Hadamard gate, $S$ is a phase gate, and $T$ is the $\pi/4$ phase gate. The entire gate sequence is $ABBBAAAABBABABABBBAA$. These gates are simulated following the rules of fault tolerance as described in \cite{YSWTgate,How,How2}. Clifford gates are implemented bit-wise. $T$-gates are implemented via an ancilla state $|\Theta\rangle = \frac{1}{\sqrt{2}}(|0_L\rangle+e^{i\frac{\pi}{4}}|1_L\rangle)$, where $|0_L\rangle$ and $|1_L\rangle$ are the logical basis states of the [[7,1,3]] QEC code. A bit-wise CNOT gate is applied between the qubits of the state $|\Theta\rangle$ and the qubits of the encoded state with the $|\Theta\rangle$ state qubits as control. The encoded state is measured, and if the outcome is zero the qubits of the $|\Theta\rangle$ state will be projected into the state $T$ acting on the encoded state. We stress that the entire process is simulated in the non-equiprobable error environment including the logical gates themselves, initialization, measurement, and SM including construction of the Shor and Steane states and all verifications. In these simulations, we assume that no error is detected for any of the 50 QEC applications and we start with the initial logical state $|0\rangle$. Other initial states give similar results. 

We denote the final state after application of the 50 gates in the noisy environment as $\rho_f$ and determine the accuracy of this state via the fidelity between $\rho_f$ and the ideal final state if there were no errors, $\rho_i$. The fidelity is given by $F = \rm{Tr}[\rho_f\rho_i]$. We will find it convenient to utilize the logarithmic infidelity $-\log_{10}(1-F)$.   

\begin{figure}
\includegraphics[width=6cm]{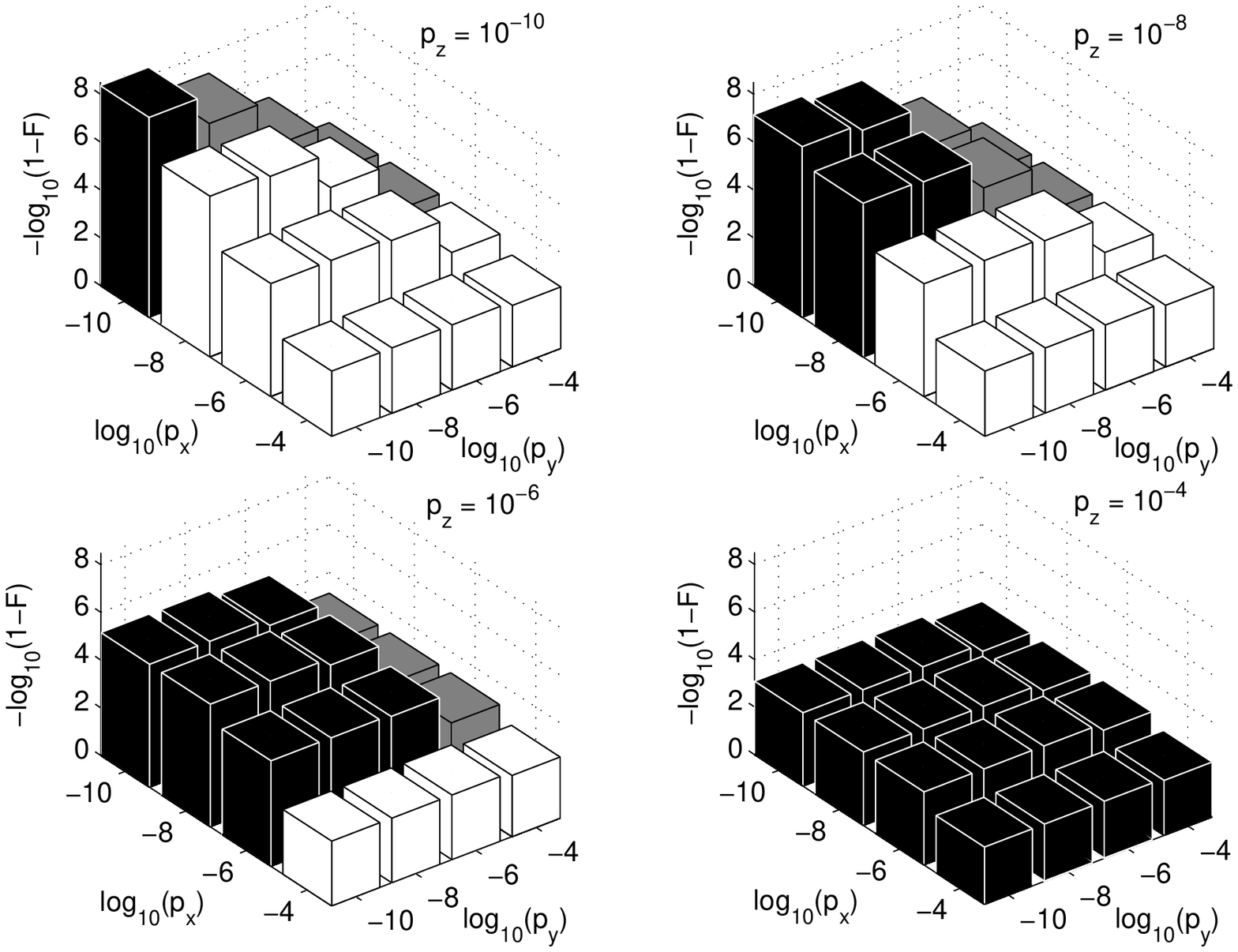}
\includegraphics[width=6cm]{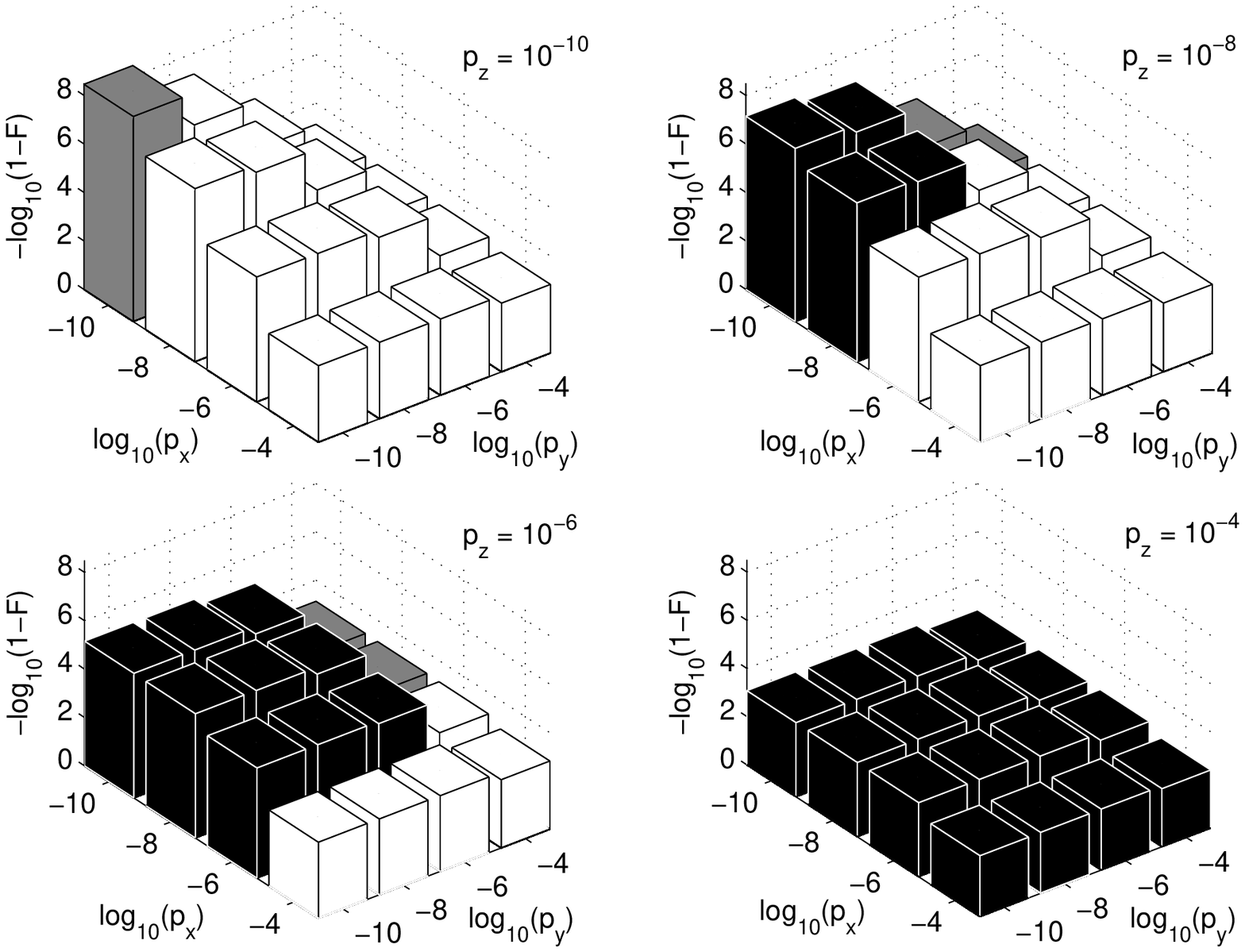}
\caption{Logarithmic infidelity of the final state after 50 logical gates for Shor state SM (left) and Steane state SM (right). The results are shown for 16 different error environments parameterized by the different error probabilities for each error type. The color shows which syndrome order gives the output state with the highest fidelity for that particular error environment (black for $ZXXZ$, gray for $XZXZ$, and white for $ZXZX$).   
}
\label{Fifty}
\end{figure}

\begin{figure}
\includegraphics[width=6cm]{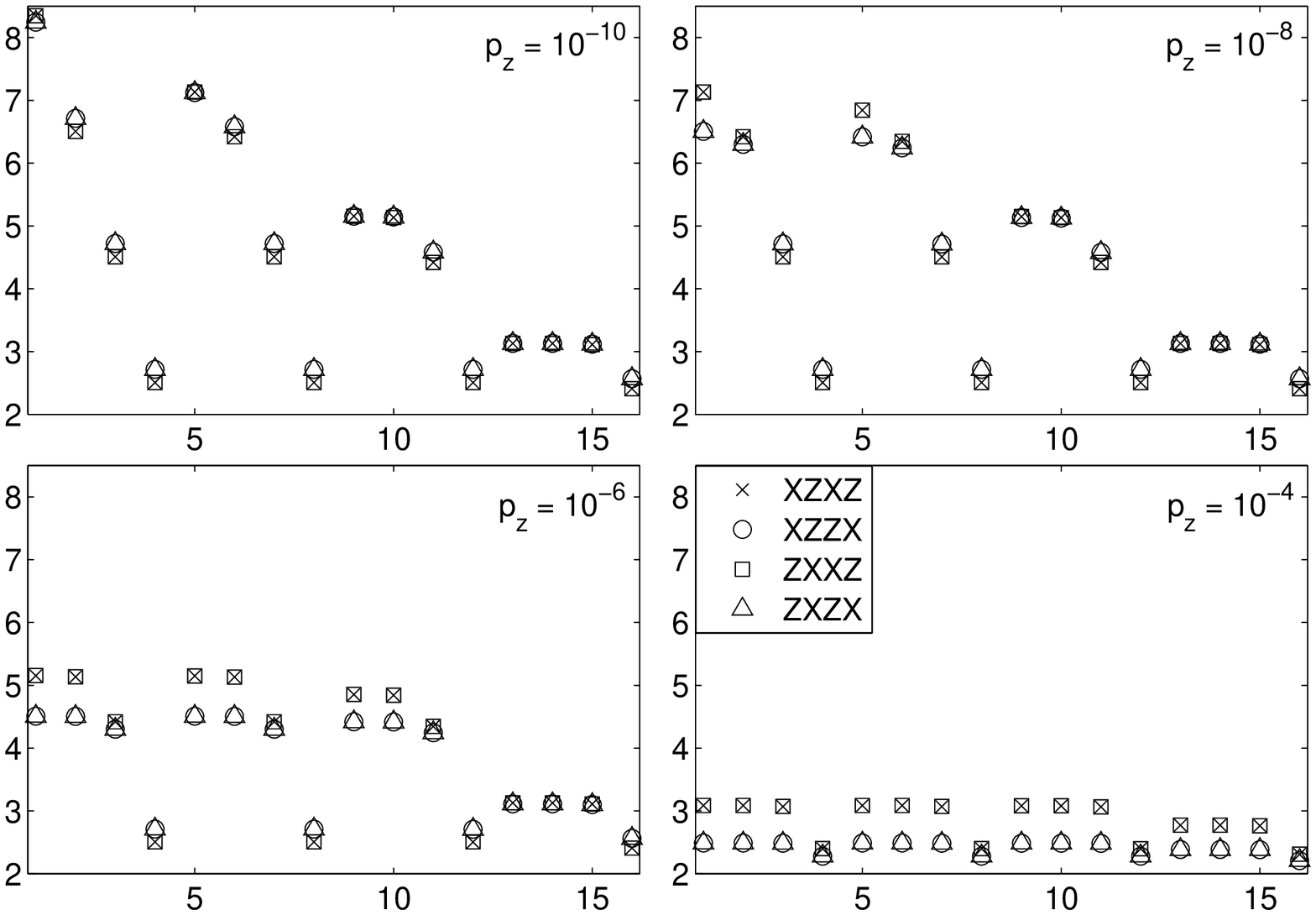}
\includegraphics[width=6cm]{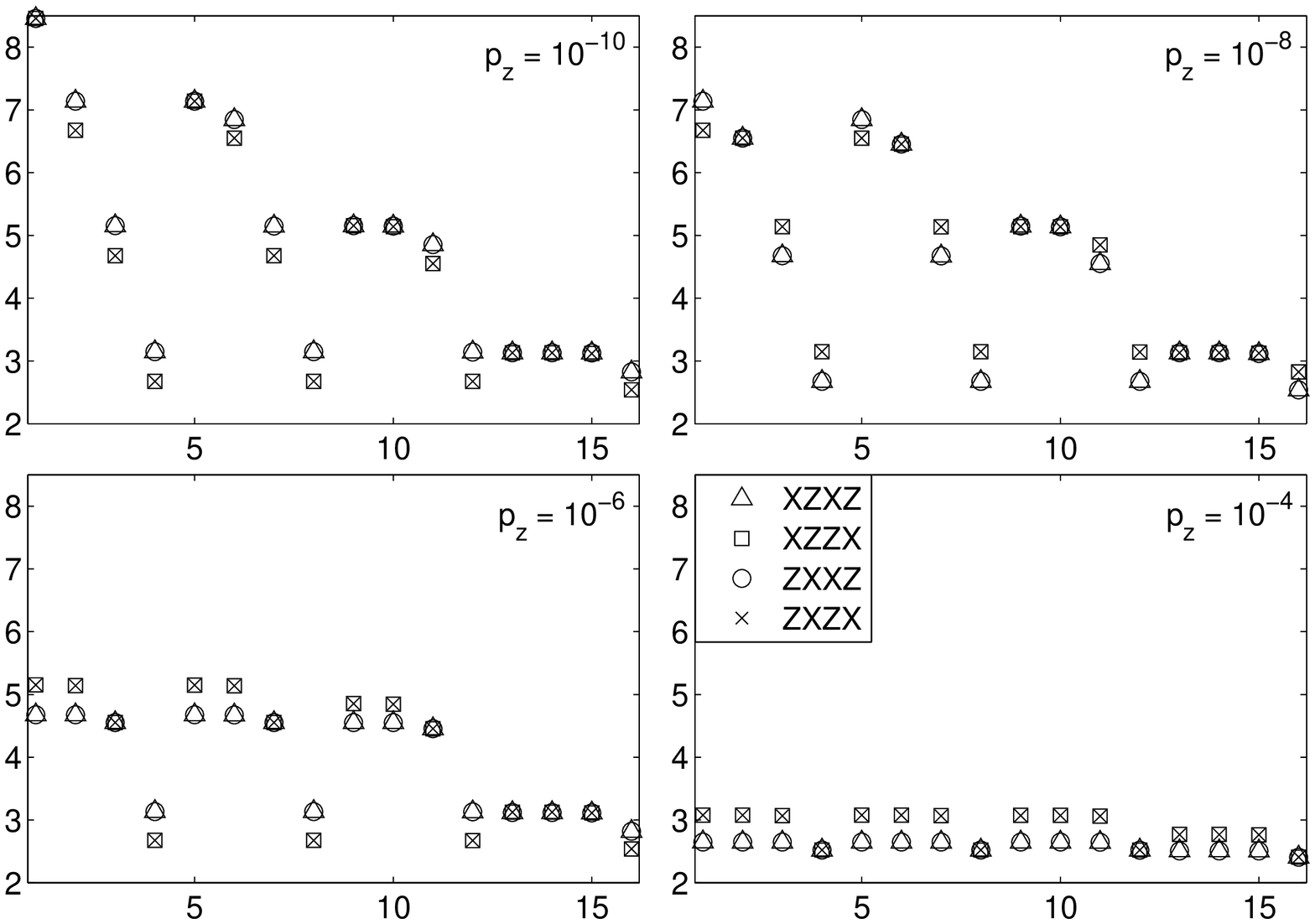}
\caption{Logarithmic infidelity of the final state after 50 logical gates for Shor state SM (left) and Steane state SM (right). The results shown are for all four SM orders. The x-axis of each enumerates the different error environments. The first element is for an environment where $p_x = p_y = 10^{-10}$. The next three elements are for increased vaules of $p_x$ in steps of two orders of magnitude. The fifth element resets $p_x$ to $10^{-10}$ and increases $p_y$ by two orders of manitude. This continues for the 16 elements in each subplot. Easily noticeable is the similarity between the pairs of syndrome orders ending with either $XZ$ or $ZX$. Also important is that the cost of using the wrong syndrome `order' can be above .5 in logarithmic infidelity.   
}
\label{Fifty2}
\end{figure}

Figures \ref{Fifty} and \ref{Fifty2} show the results of our simulations. Figure \ref{Fifty} graphs the logarithmic infidelity of the most accurate of the four final states (identified by the color of the bar) as a function of the single-qubit error probabilities. When utilizing Shor state SM or Steane state SM, for most of the error environments, the syndrome order $ZXXZ$ gives the highest fidelities especially when phase flip errors are dominant or in a depolarizing environment. When bit flip errors are dominant the syndrome order $ZXZX$ tends to give the highest fidelities. And, when $\sigma_y$ errors are dominant the syndrome order $XZXZ$ gives the highest fidelity. Figure \ref{Fifty2} compares the logarithmic infidelities for the different orders. It is clearly seen that the difference between the best fidelity and second best fidelity tends to be slight, especially between the syndrome order pairs $ZXXZ$ and $XZXZ$, and $ZXZX$ and $XZZX$. Thus, while for some error environments choosing a suboptimal syndrome order may lead to a difference in logarithmic infidelity of up to 0.65, for other suboptimal choices the difference is much less. 

The above results demonstrate that, in general, a proper strategy is to apply the SM of the most dominant error last. This is not surprising. If the SM for the non-dominant error was applied last there would be more time for the dominant error to work uncorrected. This point is substantiated by the grouping of pairs by the last SM applied. The grouping also shows the lack of influence of the first bit-flip and phase-flip SM set. Once the second set is implemented the dependence on order of the first is wiped out. Based on this we can fairly generalize our results to cases when more than two rounds of SM are necessary. The last set of bit-flip and phase-flip SM seems to determine the proper SM order. Despite these generalities, there is subtletly in these results especially with the introduction of $\sigma_y$ errors. This is because both the bit-flip and phase-flip SM are necessary to identify and locate with type of error. In additoin, certain SM may be more sensitive to certain types of error \cite{WB}. The results of all of these simulations is to allow a quantum computer programmer, knowing the error environment of the quantum computer, to choose the syndrome order that will yield the highest fidelity. 

\begin{figure}
\begin{center}
\includegraphics[width=6cm]{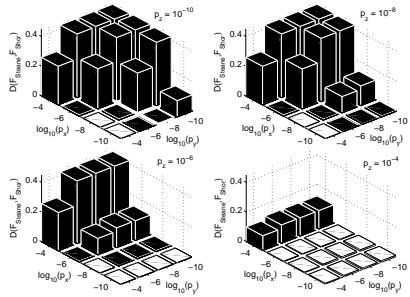}
\caption{Absolute values of the difference in logarithmic infidelities, $D(F_{Steane},F_{Shor})$, between final states with Steane state SM and Shor state SM as a function of error probabilities. Black bars indicate environments in which Steane state SM give higher fidelities and white bars indicate when Shor state SM give higher fidelities. 
}
\label{comp}
\end{center}
\end{figure}

We now compare the fidelities of the output states depending on whether Shor and Steane states were used to implement SM. While not the primary aim of this work (see \cite{YSWSynds} for a more extensive treatment), this comparison is important for those looking to optimize applied SM for particular systems. Fig.~\ref{comp} shows $D(F_{Steane},F_{Shor})$, which we define as the absolute value of the difference in logarithmic infidelity between the final states utilizing Steane state and Shor state SM, as a function of the different error probabilities. In general, using Steane states leads to higher fidelities. However, in phase flip or $\sigma_y$ error dominant environments, the Shor state SM provide a slightly higher fidelity. This is in consonance with previous work demonstrating that Shor state SM is particularly sensitive to bit-flip errors \cite{WB}.     

Some of the general results of this study should not be surprising. It makes sense to apply as your final SM the one that will correct the most dominant error. Identifying when this is not the case, especially when $\sigma_y$ errors are a factor, is necessary and demonstrates one of the important aspects of this work. The fact that the order of the initial set of SM is almost irrelevant is also interesting and again emphasizes the importance of the final step of a given quantum protocol. The comparison of the Shor and Steane state SM will be useful for optimizing SM in a given error environment.   

This study may also be useful in that it demonstrates the utility of fidelity in testing the accuracy of quantum protocols. Studies of QFT generally rely on determining logical error probabilities. This, in turn, is based on the number of possible ways a logical qubit can have an error. All of our simulations were done following the rules of QFT. In addition, for a given SM type the number of possible ways a logical qubit can have an error is independent of the SM order. Nevertheless, the fidelity can be very different. We believe that the fidelity may be useful even when a protocol does not follow the tenets of QFT \cite{YSWTgate,YSW}. This is because the fidelity does give the accuracy of the final state. Even if there is a possibility of uncorrectable errors to logical qubits, this can be circumvented by repeating the algorithm multiple times. If the fidelity is high we can be almost certain that a reasonable number of repetitions will eventually yield the correct output. 

This article is dedicated in memory of Dr.~Howard Brandt. Those of use fortunate enough to know Howard and have the opportunity to interact with him know that he had a heart of gold. However, one thing that Howard did not react well to was bad science. Howard had no qualms in calling out those who he felt were not accurate or were simply wrong. It may not have endeared him to those he called out, but it was the mark of a good scientist and a good Editor. I am honored to be able to contribute to the Special Issue in his memory. 


\begin{acknowledgements}
This research is supported under the MITRE Innovation Program. \copyright 2015 The MITRE Corporation. ALL RIGHTS RESERVED. Approved for Public Release; Distribution Unlimited. Case Number 15-1438.
\end{acknowledgements}

\end{document}